\documentclass[12pt,a4paper]{article}

\usepackage{amsmath,amssymb}
\usepackage{bm}
\usepackage{graphicx}
\usepackage{ascmac}
\usepackage{wrapfig}
\usepackage{braket}
\usepackage{color}
\usepackage{mathrsfs}
\usepackage[english]{babel}
\usepackage{inconsolata}
\usepackage{cases}
\usepackage{tikz}
\usepackage{cite}
\usepackage{setspace}
\usepackage[top=3cm, bottom=1.8cm, left=2.3cm, right=2.3cm, includefoot]{geometry}
\usepackage{hyperref}

\usepackage{array}
\usepackage{longtable}
\usepackage{colortab}
\usepackage{colortbl}
\usepackage{arydshln}

\newcommand{\beq}{\begin{equation}}
\newcommand{\eeq}{\end{equation}}
\newcommand{\lr}[1]{\left(#1\right)}
\newcommand{\lrk}[1]{\left[#1\right]}
\newcommand{\Schro}{Schr$\rm{\ddot{o}}$dinger\ }

\makeatletter
\@addtoreset{equation}{section}

\@addtoreset{figure}{section}

\@addtoreset{table}{section}

\makeatother

\begin{document}

\begin{titlepage}

\pagenumbering{gobble}

\begin{flushright}
\rm TIT/HEP-692 \\ December, 2022
\end{flushright}

\renewcommand{\thefootnote}{*}

\vspace{0.2in}
\begin{center}
\Large{\bf Exact conditions for Quasi-normal modes of \\ extremal M5-branes and Exact WKB analysis}
\end{center}
\vspace{0.2in}
\begin{center}
\large Keita Imaizumi\footnote{E-mail: k.imaizumi@th.phys.titech.ac.jp}
\end{center}

\begin{center}{\it
Department of Physics,
\par
Tokyo Institute of Technology
\par
Tokyo, 152-8551, Japan
}
\end{center}
\vspace{0.2in}
\begin{abstract}
\begin{spacing}{1.0}
{\footnotesize  
We study the quasi-normal modes (QNMs) of a massless scalar perturbation to the extremal M5-branes metric by using the exact WKB analysis. The exact WKB analysis provides two exact QNMs conditions depending on the argument of the complex frequency of the perturbation. The exact conditions show that the discontinuity of the perturbative part of the QNMs leads the non-perturbative part of themselves. We also find a new example of the Seiberg-Witten/gravity correspondence, which helps us to compute the QNMs from our conditions.}
\end{spacing}
\end{abstract}

\end{titlepage}

\newpage

\pagenumbering{arabic}
\setcounter{page}{1}

\renewcommand{\thefootnote}{\arabic{footnote}}
\setcounter{footnote}{0}

\section{Introduction}
\label{sec:intro}
The exact WKB analysis for the complex one-dimensional stationary \Schro equation is an exact approach to study the spectral problems \cite{Vor, DP, cr1}. In this formulation, we can derive exact energy quantization conditions, which include the non-perturbative contributions arising from the quantum tunneling effect, through the connection formula of the wavefunction (e.g. \cite{Esemi, 1811.04812, 2008.00379, 2008.13680}). The exact quantization conditions not only determine the energy eigenvalues exactly, but also clarify that the non-perturbative parts of the energy eigenvalues can be captured from the discontinuity of the perturbative parts of themselves. 
\par
Recently, the exact WKB analysis has been applied to study eigenfrequencies of a perturbation to a black hole metric \cite{2207.09961}, where we can compute the complex resonant eigenfrequencies of the perturbation called the quasi-normal modes (QNMs). In \cite{2207.09961}, we derived an exact condition for the QNMs of the massless dilaton perturbation to the extremal D3-brane metric by using the exact WKB analysis. The QNMs computed by the exact condition were consistent with the Leaver's numerical method \cite{Leaver}, and the result of the classical WKB approximation at large frequency \cite{classicalWKB}. This shows that the exact WKB analysis is a powerful tool to study the QNMs spectral problems.
\par
There is another approach for computing the QNMs, based on the Seiberg-Witten theory for $\mathcal{N}=2$ supersymmetric gauge theory. In \cite{2006.06111}, the authors shown that the E.O.M. for perturbations to the 4-dimensional Schwarzschild black hole and the Kerr black hole can be transformed to the quantum Seiberg-Witten curve for 4-dimensional $\mathcal{N} = 2$ SU(2) supersymmetric QCD \cite{1705.09120}. This transformation enables us to utilize gauge theory techniques for solving the QNMs spectral problems. The Seiberg-Witten/gravity correspondence has extended to other geometries including D3-branes, fuzzballs and exotic compact objects later \cite{2105.04245, 2109.09804}. It is interesting to explore this correspondence in other geometries.
\par
The purpose of this paper is to study the QNMs of a massless scalar perturbation to the extremal M5-brane metric based on the exact WKB analysis. The scattering problem of the massless scalar in the extremal M5-branes spacetime was considered in \cite{hep-th/9702076}, where the author computed the classical cross section for the absorption of the massless scalar by the extremal M5-branes and compared with the result computed in the world volume effective theory of the M5-branes. In the present paper, we compute the exact QNMs conditions for the massless scalar perturbation to the extremal M5-branes. We see that the exact WKB analysis provides two exact QNMs conditions depending on the argument of the complex frequency of the perturbation. The exact conditions not only determine the QNMs non-perturbatively, but also show that the discontinuity of the perturbative part of the QNMs leads the non-perturbative part of themselves. We analytically and numerically compute the QNMs by using the exact conditions. We also show that the E.O.M. for the massless scalar perturbation can be transformed to the quantum Seiberg-Witten curve for 4-dimensional $\mathcal{N} = 2$ SU(2) supersymmetric QCD with a massless fundamental matter \cite{1705.09120}. This transformation will be useful to compute the QNMs.
\par
This paper is organized as follows. In section \ref{sec:EWKB}, we introduce the E.O.M. for the massless scalar perturbation to the extremal M5-brane metric and the exact WKB analysis. We review the WKB method to the one-dimensional stationary \Schro equation and the Borel resummation, which promotes the WKB method to the exact WKB analysis. In section \ref{sec:QSP}, we solve the QNMs spectral problem for the massless scalar perturbation to the extremal M5-brane metric by using the exact WKB analysis. Combining the boundary conditions, we derive the exact conditions for the QNMs. In section \ref{sec:CompQNM}, we compute the QNMs based on our exact conditions.

\section{Exact WKB analysis for extremal M5-branes}
\label{sec:EWKB}

\subsection{E.O.M. for scalar field in extremal M5-branes metric}
\label{subsec:EOMM5}

The geometry of a stack of the extremal M5-branes is given by the following line element \cite{M5metric},
\beq
\label{eq:M5metric}
 ds^2 = A(r)^{-1/3}\lr{-dt^2 + dx_1^2 + \cdots + dx_5^2} + A(r)^{2/3}\lr{dr^2 + r^2d\Omega_4^2},
\eeq
where $A(r) = 1 + \frac{R^3}{r^3}$, $R$ is a positive real parameter, $\lr{t, x_1, \cdots, x_5}$ are the coordinates on the M5-branes world-volume, $r$ is the radial coordinate and $d\Omega_4^2$ is the metric on $S^4$ in the bulk of the M5-branes. We consider a massless scalar field $\Phi$ in this background obeying the E.O.M.:
\beq
\label{eq:EOMM5}
 \Box\Phi = \frac{1}{\sqrt{-g}}\partial_{M}\lr{\sqrt{-g}g^{MN}\partial_{N}}\Phi = 0,
\eeq
where $g_{MN}\ \lr{M, N = 0, 1, \cdots, 10}$ is the metric given by (\ref{eq:M5metric}) and $g = \det g_{MN}$. Under the separation of the variables, the solution to (\ref{eq:EOMM5}) takes the form
\beq
\Phi = e^{-iEt + i\bm{k}\cdot\bm{x}}r^4\phi(r)\mathcal{Y}_{\bm{m}}^{l}(\bm{\theta}),
\eeq
where $\mathcal{Y}_{\bm{m}}^{l}(\bm{\theta})$ is the spherical harmonics on $S^4$ obeying $\Delta_{S^4}\mathcal{Y}_{\bm{m}}^{l}(\bm{\theta}) = -l\lr{l+3}\mathcal{Y}_{\bm{m}}^{l}(\bm{\theta})$, $l$ and $\bm{m} = (m_1, m_2, m_3)$ are integers satisfying $|m_1| \leq m_2 \leq m_3 \leq l$ and $\bm{\theta}$ denotes the angular coordinates on $S^4$. The radial part $\phi(r)$ of $\Phi$ satisfies the second-order differential equation \cite{hep-th/9702076},
\beq
\label{eq:radM5}
 \lrk{\rho^{-4}\frac{d}{d\rho}\rho^4\frac{d}{d\rho} + 1 + \frac{(\omega R)^3}{\rho^3} - \frac{l\lr{l+3}}{\rho^2}}\phi(\rho/\omega) = 0,\ \ (\rho = \omega r)
\eeq
where $\omega = \sqrt{E^2 - \bm{k}^2}$. Introducing $z$ by $\rho = \omega R z$ and $\psi(z)$ by $\phi(\rho/\omega) = (\omega R z)^{-2}\psi(z)$, (\ref{eq:radM5}) becomes the \Schro type differential equation,
\beq
\label{eq:Schro}
 \lrk{\frac{d^2}{dz^2} -\eta^2Q_0(z) - Q_2(z)}\psi(z) = 0,
\eeq
with
\beq
 Q_0(z) = -\frac{\lr{\omega R}^2\lr{z^3 + 1} - \lr{l+\frac{3}{2}}^2 z}{z^3}, \ \ \ Q_2(z) = -\frac{1}{4z^2},
\eeq
and $\eta = 1$. We keep $\eta$ as a complex parameter, which plays a role of the Planck constant $\hbar = 1/\eta$ \footnote{This notation follows \cite{cr1}}.
\par
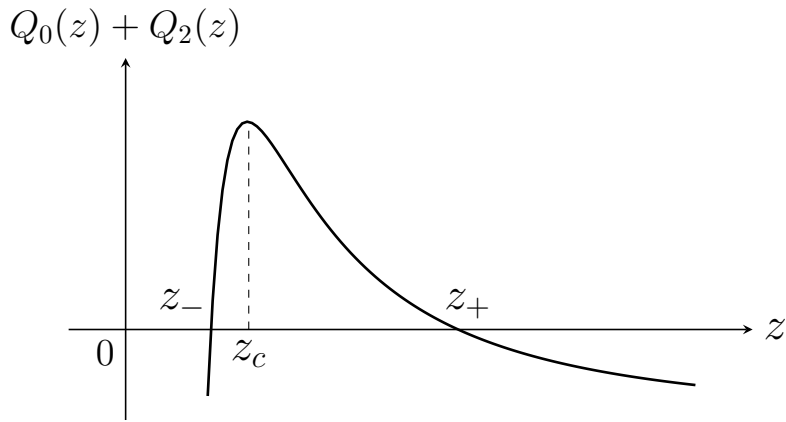
\begin{figure}[t]
  \begin{center}
  \hspace*{0.3in}
   \begin{tikzpicture}[xscale=3, yscale=1.2]
    \draw[->,>=stealth,semithick] (-0.25,0)--(2.75,0) node[right]{\Large$z$};
    \draw[->,>=stealth,semithick] (0,-1.0)--(0,3) node[above]{\large$Q_0(z) + Q_2(z)$};
    \draw (0,0)node[below left]{\large$0$};
    \draw[samples=100,domain=0.36:2.5,line width=1pt] plot(\x,{-(1)^2*(1 + 1/((\x)*(\x)*(\x))) + (1/4 + 3/2)^2/((\x)*(\x)) - 1/(4*(\x)*(\x))});
    
    \draw (0.40,0)node[above left]{\Large$z_-$};
    \draw (1.35,0)node[above right]{\Large$z_+$};
    
    \draw[dashed] (0.54,0)--(0.54,2.25);
    \draw (0.54,0)node[below]{\Large$z_c$};
   \end{tikzpicture}
   \caption{A potential graph of $Q_0(z) + Q_2(z)$.}
   \label{fig:Q0Q2}
  \end{center}
\end{figure}

We will consider the solution satisfying the following boundary conditions: (i) only outgoing wave exists at $z \rightarrow +\infty$ along the real axis and (ii) only ingoing wave exists at $z \rightarrow 0$. Then we obtain the set of the spectrum of $\omega$ called the QNMs, which take discrete values with positive real and negative imaginary part. We first study the semiclassical value of the QNMs for large $l$, at which the QNMs can be determined from the Bohr-Sommerfeld quantization condition \cite{classicalWKB},
\beq
\label{eq:largeO}
 \int_{z_-}^{z_+}\sqrt{Q_0(z) + Q_2(z)}dz = \pi i\lr{n + \frac{1}{2}},
\eeq
where $n \in \mathbb{Z}_{\geq 0}$ and $z_{\pm}$ are the turning points satisfying $Q_0(z_{\pm}) + Q_2(z_{\pm}) = 0$ (Fig.\ref{fig:Q0Q2}). At large $l$, the turning points close each other and the l.h.s. of (\ref{eq:largeO}) can be integrated by approximating $Q_0(z) + Q_2(z)$ by its expansion at the maximum point  $z_c$ (Fig.\ref{fig:Q0Q2}). Then (\ref{eq:largeO}) reduces to
\beq
\label{eq:largeOapp}
 \frac{Q_0(z_c) + Q_2(z_c)}{\sqrt{2\lr{Q_0''(z_c) + Q_2''(z_c)}}} = i\lr{n + \frac{1}{2}}.
\eeq
The condition (\ref{eq:largeOapp}) can be solved perturbatively with respect to the imaginary part of $\omega R$. Up to the linear order, one finds
\beq
\label{eq:largeOappAF}
 \omega R = \frac{\sqrt[3]{2}}{\sqrt{3}}\sqrt{\lr{l + \frac{3}{2}}^2-\frac{1}{4}} -i\frac{\sqrt[3]{2}}{\sqrt{3}}\lr{n+\frac{1}{2}}.
\eeq
The formula (\ref{eq:largeOappAF}) indicates that the real part of the QNMs only depends on $l$ and the imaginary part only depends on $n$ in the large $l$ region.
\par
The Bohr-Sommerfeld quantization condition is only valid for large $l$. One of the way to compute the QNMs even at small $l$ is to determine exact conditions for the QNMs by using the exact WKB analysis \cite{2207.09961}. In this paper, we study the spectral problem of (\ref{eq:Schro}) based on the Exact WKB analysis and derive exact conditions for the QNMs of the massless scalar in the extremal M5-branes background.

\subsection{Extremal M5-branes and WKB method}
\label{subsec:OWKB}
We study the solution to (\ref{eq:Schro}) by using the exact WKB analysis. The coordinate $z$ is regarded as a complex variable. We consider the following ansatz of the solution,
\beq
\label{eq:WKBsol}
 \psi_a(z) = \exp{\left[\eta\int_a^zP(z')dz'\right]},
\eeq
where $a$ is a parameter controlling the normalization of the solution. Substituting (\ref{eq:WKBsol}) into (\ref{eq:Schro}), we obtain the Riccati equation for $P(z)$,
\beq
\label{eq:Ricatti}
 P^2(z) + \eta^{-1}\frac{d}{dz}P(z) - Q_0(z) - \eta^{-2}Q_2(z) = 0.
\eeq
(\ref{eq:Ricatti}) can be solved by expanding $P(z)$ as a power series in $\eta^{-1}$,
\beq
\label{eq:Pexp}
 P(z) = \sum_{n=0}^{\infty}p_n(z)\eta^{-n}.
\eeq
Plugging (\ref{eq:Pexp}) in (\ref{eq:Ricatti}) and equating the coefficients of the powers of $\eta^{-1}$, $p_n(z)$ is determined recursively,
\beq
\label{eq:pnrec}
\begin{split}
 & p_0^2(z) = Q_0(z), \\
 & 2p_0(z)p_1(z) + \frac{d}{dz}p_0(z) = 0, \\
 & 2p_0(z)p_2(z) + p_1^2(z) + \frac{d}{dz}p_1(z) = Q_2(z), \\
 & 2p_0(z)p_n(z) + \sum_{i, j \geq 1, i + j = n}p_i(z)p_j(z) + \frac{d}{dz}p_{n-1}(z) = 0, \ \ (n \geq 3). 
\end{split}
\eeq
Depending on the choice of the sign for the root of the first equation in (\ref{eq:pnrec}), we obtain two functions $\pm P(z)$ with $p_0(z) = \sqrt{Q_0(z)}$. We denote the solution for each sign as
\beq
\label{eq:pmWKB}
 \psi_a^{\pm}(z) = \exp{\left[\pm\eta\int_a^zP(z')dz'\right]}.
\eeq
\\
If we split (\ref{eq:Pexp}) into even and odd powers of $\eta^{-1}$ as,
\beq
 P(z) = \sum_{n=0}^{\infty}p_{2n}(z)\eta^{-2n} + \sum_{n=0}^{\infty}p_{2n+1}(z)\eta^{-(2n+1)} = P_{\mathrm{even}}(z) + P_{\mathrm{odd}}(z),
\eeq
the odd power collection of (\ref{eq:Ricatti}) shows that $P_{\mathrm{odd}}(z)$ is a total derivative, 
\beq
 P_{\mathrm{odd}}(z) = -\frac{\eta^{-1}}{2P_{\mathrm{even}}(z)}\frac{d}{dz}P_{\mathrm{even}}(z) = -\frac{\eta^{-1}}{2}\frac{d}{dz}\log{P_{\mathrm{even}}(z)}.
\eeq
Therefore the solutions (\ref{eq:pmWKB}) can be written by only using $P_{\mathrm{even}}(z)$ as,
\beq
\label{eq:evenWKB}
 \psi_a^{\pm}(z) = \frac{1}{\sqrt{P_{\mathrm{even}}(z)}}\exp{\left[\pm\eta\int_a^zP_{\mathrm{even}}(z')dz'\right]}.
\eeq
In this paper, we refer the solutions (\ref{eq:evenWKB}) as WKB solutions. The WKB solutions can be expanded as the series in $\eta^{-1}$,
\beq
\label{eq:evenWKBexp}
 \psi_a^{\pm}(z) = \exp(\pm\eta\int_{a}^{z}p_0(z')dz')\sum_{n=0}^{\infty}\psi_{a, n}^{\pm}\lr{z}\eta^{-(n+\frac{1}{2})}.
\eeq
\par
$P_{\mathrm{even}}(z)dz$ in (\ref{eq:evenWKB}) can be regarded as a one-form on the Riemann surface $\Sigma$ defined by the following algebraic curve,
\beq
 \Sigma : y^2 = Q_0(z),
\eeq
where the branch points on $\Sigma$ are determined by the zeros of $Q_0(z)$ and $z = \infty$. $\Sigma$ is a double covering for the complex plane. If we move from a sheet to the other, the sing of $\pm P_{\mathrm{even}}(z)dz$ is reversed and the WKB solutions are also changed $\psi_a^{\pm}(z) \rightarrow \psi_a^{\mp}(z)$. The one-cycle $\gamma \in H_1(\Sigma)$ generates the period of $P_{\mathrm{even}}(z)dz$,
\beq
\label{eq:period}
 \Pi_{\gamma} := \oint_{\gamma}P_{\mathrm{even}}(z)dz, \ \ \ \gamma \in H_1(\Sigma).
\eeq
We refer (\ref{eq:period}) as WKB period. The WKB periods are even power series in $\eta^{-1}$ as $P_{\mathrm{even}}(z)$,
\beq
\label{eq:periodexp}
 \Pi_{\gamma} = \sum_{n=0}^{\infty}\Pi_{\gamma}^{(2n)}\eta^{-2n}, \ \ \Pi_{\gamma}^{(2n)} = \oint_{\gamma}p_{2n}(z)dz.
\eeq

\subsection{Borel resummation}
\label{subsec:Bsum}
The WKB solutions and the WKB periods are asymptotic series converging only at $\eta = \infty$. In the exact WKB analysis, they are promoted to analytic functions by Borel resummation. Let us consider a general asymptotic series $f$,
\beq
 f = e^{-\eta A}\sum_{n=0}^{\infty}a_n\eta^{-(n+\alpha)},\ \ \lr{\alpha \notin \{-1, -2, -3, \cdots\}}.
\eeq
The Borel transformation of $f$ is defined as,
\beq
\label{eq:BT}
 \hat{f}\lr{\xi} = \sum_{n=0}^{\infty}\frac{a_n}{\Gamma\lr{n+\alpha}}\lr{\xi - A}^{n + \alpha -1}.
\eeq
If (\ref{eq:BT}) converges near $\xi = A$ and can be analytically continued  on the complex $\xi$-plane, $f$ is said to be Borel summable (definition 2.10 of \cite{cr1}). The Borel resummation of $f$ is defined as the following integral of the Borel transformation,
\beq
\label{eq:Bsumdef}
 \mathcal{B}\lrk{f} = \int_{A}^{\infty e^{-i\theta}}e^{-\eta\xi}\hat{f}\lr{\xi}d\xi, \ \ \ (\theta = \arg(\eta)).
\eeq
The Borel resummaton is analytic on $\eta$ and has $f$ as the asymptotic expansion at $\eta = \infty$. For any Borel summable series $f, g$, the Borel resummation commutes with addition and multiplication (proposition 2.11 of \cite{cr1}),
\beq
\label{eq:Bsumalge}
\begin{split}
 \mathcal{B}\lrk{f + g} &= \mathcal{B}\lrk{f} +\mathcal{B}\lrk{g}, \\
 \mathcal{B}\lrk{fg} &= \mathcal{B}\lrk{f}\mathcal{B}\lrk{g}.
\end{split}
\eeq
\par
The Borel resummed WKB solutions $\mathcal{B}\lrk{\psi_a^{\pm}}(z)$ are only local solutions at $z$. To obtain the solutions in the whole of the complex plane, we need to analytically continue Borel resummed WKB solutions. The global structure of the Borel resummed WKB solutions is determined by the Stokes lines, which are defined by the following equation,
\beq
\label{eq:SL}
 \Im{\lrk{e^{i\theta}\int_{z_*}^{z}p_0\lr{z'}dz'}} = 0,
\eeq
where $z_{*}$ is a zero of $Q_0(z)$. The endpoints of the Stokes lines are on $z_*$ or $z = 0, \infty$, which are the irregular singular points of the differential equation (\ref{eq:Schro}). The Stokes lines can be assigned the orientation defined by the following conditions,
\beq
\label{eq:Orientation}
\left\{
\begin{array}{l}
\Re\lrk{e^{i\theta}\int_{z_*}^{z}p_0\lr{z'}dz'} > 0 \rightarrow \mathrm{positive\ orientation}, \\
\Re\lrk{e^{i\theta}\int_{z_*}^{z}p_0\lr{z'}dz'} < 0 \rightarrow \mathrm{negative\ orientation}.
\end{array}
\right.
\eeq
The graph on $\Sigma$ depicted by the Stokes lines is called Stokes graph. The Stokes graph splits $\Sigma$ into some regions, which can be mapped into a triangulation of $\Sigma$ \cite{cr1, 0907.3987}. Each region divided by the Stokes graph is called Stokes region.
\par
The Borel resummed WKB solutions are analytic in each Stokes region. If the path of the analytic continuation crosses any stokes line, then the solutions are discontinuously changed.

\section{Spectral problem of extremal M5-branes}
\label{sec:QSP}
We will solve the connection problem of the Borel resummed WKB solutions. The solution describing the outgoing wave at the infinity is connected to both of the solutions describing ingoing and outgoing waves at the origin. The exact QNMs conditions will be given as the conditions that the outgoing wave at the origin vanishes.

\subsection{Stokes graph for extremal M5-branes}
\label{subsec:SG}
First, we discuss generic properties of the Stokes graphs of the model. Let $z_1, z_2, z_3$ be the zeros of $Q_0(z)$ ordered as $\Re(z_1) \leq \Re(z_2) \leq\Re(z_3)$. We assume $z_1, z_2, z_3$ are simple zeros. The  three Stokes lines then emanate from each zero. The endpoints of the Stokes lines are on either $z_1, z_2, z_3, 0, \infty$. By substituting the explicit form of $p_0(z) = \sqrt{Q_0(z)}$ into the definition of the Stokes lines (\ref{eq:SL}) and dividing by the real parameter $l+\frac{3}{2}$, one finds that (\ref{eq:SL}) depends on only the combination of the parameters $\omega R /\lr{l + \frac{3}{2}}$ and $\eta$. 
\par
The Stokes graphs can be classified according to how the Stokes regions are adjacent to each other (or which triangulation of $\Sigma$ the Stokes graph is mapped). If we continuously change $\theta = \arg(\eta)$, the state of the adjacent of the Stokes regions may be discontinuously changed at specific values of $\arg(\eta)$. We refer the discontinuous change of the Stokes graph as saddle reduction \cite{cr1}. We specify when this saddle reduction occurs with respect to the value of $\omega R /\lr{l + \frac{3}{2}}$ instead of $\arg(\eta)$. The sign of the saddle reduction is to appear the Stokes line that connects two zeros of $Q_0(z)$. By definition of the Stokes lines (\ref{eq:SL}), the parameters at which two of $z_1, z_2, z_3$ are connected can be determined by the following equations,

\begin{figure}[t]
  \begin{center}
   \hspace*{-0.5in}
    \includegraphics[clip, width=10.0cm]{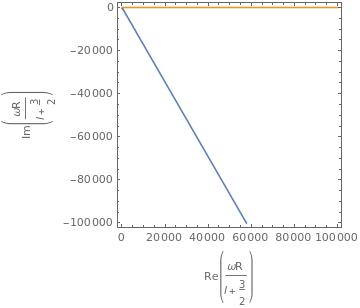}
    \caption{The parameters satisfying (\ref{eq:13cond})$\sim$(\ref{eq:12cond}) at $\arg(\eta) = 0$. The orange line indicates the value of $\omega R/\lr{l+\frac{3}{2}}$ satisfying (\ref{eq:23cond}). The blue line, which is drawn on $\arg(\omega R) = -\frac{\pi}{3}$, indicates the value satisfying (\ref{eq:12cond}). There are no points satisfying (\ref{eq:13cond}).}
    \label{fig:RegionPlot}
  \end{center}
\end{figure}

\beq
\label{eq:13cond}
 \Im\lrk{e^{i\theta}\int_{z_1}^{z_3}p_0\lr{z}dz} = 0,
\eeq
\beq
\label{eq:23cond}
 \Im\lrk{e^{i\theta}\int_{z_2}^{z_3}p_0\lr{z}dz} = 0,
\eeq
\beq
\label{eq:12cond}
 \Im\lrk{e^{i\theta}\int_{z_1}^{z_2}p_0\lr{z}dz} = 0.
\eeq
The integrals in the square brackets are the elliptic integrals and can also be expressed by the hypergeometric function \cite{hep-th/9605101, 2208.14031},
\beq
\label{eq:13int}
\int_{z_1}^{z_3}p_0\lr{z}dz = i\pi\left(l + \frac{3}{2}\right)\ _2F_1\lrk{-\frac{1}{6}, \frac{1}{6}, 1, \frac{27\omega^6 R^6}{4(l + \frac{3}{2})^6}},
\eeq
\beq
\label{eq:23int}
\int_{z_2}^{z_3}p_0\lr{z}dz = 
\left\{
\begin{array}{l}
-\frac{\pi\omega R \left(1-\frac{4(l + \frac{3}{2})^6}{27\omega^6 R^6}\right) }{2\sqrt[3]{2}\sqrt{3}} \ _2F_1\lrk{\frac{5}{6},\frac{5}{6},2,1-\frac{4(l + \frac{3}{2})^6}{27\omega^6 R^6}} \ \ \left(-\frac{\pi}{6} \leq \arg(\omega R) \leq 0\right), \\
\int_{z_1}^{z_3}p_0\lr{z}dz - e^{i\frac{\pi}{3}}\frac{\pi\omega R \left(1-\frac{4(l + \frac{3}{2})^6}{27\omega^6 R^6}\right) }{2\sqrt[3]{2}\sqrt{3}} \ _2F_1\lrk{\frac{5}{6},\frac{5}{6},2,1-\frac{4(l + \frac{3}{2})^6}{27\omega^6 R^6}} \\
\ \ \ \ \ \ \ \ \ \ \ \ \ \ \ \ \ \ \ \ \ \ \ \ \ \ \ \ \ \ \ \ \ \ \ \ \ \ \ \ \ \ \ \ \ \ \ \ \ \ \ \ \ \ \ \left(-\frac{\pi}{2} \leq \arg(\omega R) < -\frac{\pi}{6}\right),
\end{array}
\right.
\eeq
\beq
\label{eq:12int}
\int_{z_1}^{z_2}p_0\lr{z}dz = -\int_{z_2}^{z_3}p_0\lr{z}dz + \int_{z_1}^{z_3}p_0\lr{z}dz.
\eeq
These are local expressions but we can also compute each integral for other parameter regions by analytically continuing the hypergeometric function. By substituting (\ref{eq:13int})$\sim$(\ref{eq:12int}) into (\ref{eq:13cond})$\sim$(\ref{eq:12cond}), we can determine the values of the parameters occurring the saddle reduction. Fig.\ref{fig:RegionPlot} shows the values of $\omega R/\lr{l + \frac{3}{2}}$ satisfying (\ref{eq:13cond})$\sim$(\ref{eq:12cond}) at $\arg(\eta) = 0$ in a range. Fig.\ref{fig:RegionPlot} indicates that the Stokes line connecting $z_1$ and $z_2$ appears at $\arg(\omega R) = -\frac{\pi}{3}$. Therefore the state of the adjacent of the Stokes regions is discontinuously changed between $-\frac{\pi}{3} < \arg(\omega R) < 0$ and $-\frac{\pi}{2} \leq \arg(\omega R) < -\frac{\pi}{3}$.

\begin{figure}[t]
  \begin{center}
    \includegraphics[clip, width=10.0cm]{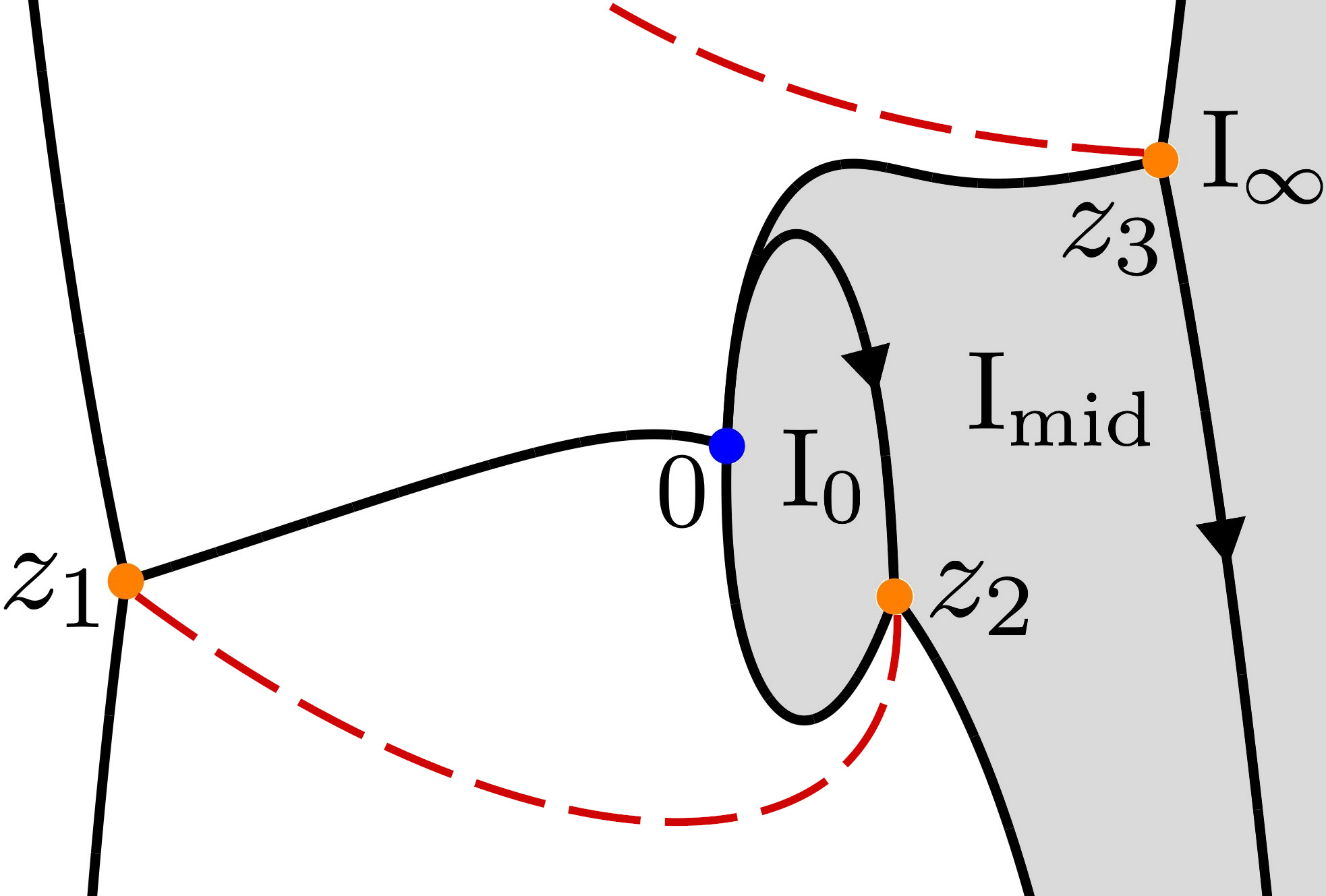}
    \caption{The Stokes graph for $\arg(\eta) = 0$ with $\omega R/\lr{l + \frac{3}{2}} = \frac{2}{3} - i\frac{2}{30}$. The sheet of $\Sigma$ is taken so that $P_{\mathrm{even}}(z)dz$ has the positive sign. The orange points $z_1, z_2, z_3$ are the zeros of $Q_0(z)$. The red dashed lines are the branch cuts (one connects $z_1$ and $z_2$, and the other connects $z_3$ and $z = \infty$). The blue point is the origin. The black lines are the Stokes lines and the black arrows show the orientation (for the negative orientation, the arrow points to the zeros of $Q_0(z)$ along the Stokes line, and for the positive orientation, the arrow points to the singular point). The stokes lines going out of the frame flow at $z = \infty$. The gray regions named $\mathrm{I}_{\mathrm{\infty}}$, $\mathrm{I}_{\mathrm{mid}}$, $\mathrm{I}_{0}$ are Stokes regions.}
    \label{fig:SGM5}
  \end{center}
\end{figure}

Let us see examples of the Stokes graphs. Fig.\ref{fig:SGM5} shows the Stokes graph for $\arg(\eta) = 0$ and $\omega R/\lr{l + \frac{3}{2}} = \frac{2}{3} - i\frac{2}{30}$, which is in the region $-\frac{\pi}{3} < \arg(\omega R) < 0$. For $\arg(\eta) = 0$ and $-\frac{\pi}{3} < \arg(\omega R) < 0$, the Stokes regions are adjacent each other in the same way as Fig.\ref{fig:SGM5} (for example, the Stokes region $\mathrm{I}_{\infty}$ is adjacent to $\mathrm{I}_{\mathrm{mid}}$, and $\mathrm{I}_{\mathrm{mid}}$ is adjacent to $\mathrm{I}_{0}$.).

\begin{figure}[t]
  \begin{center}
    \includegraphics[clip, width=12.9cm]{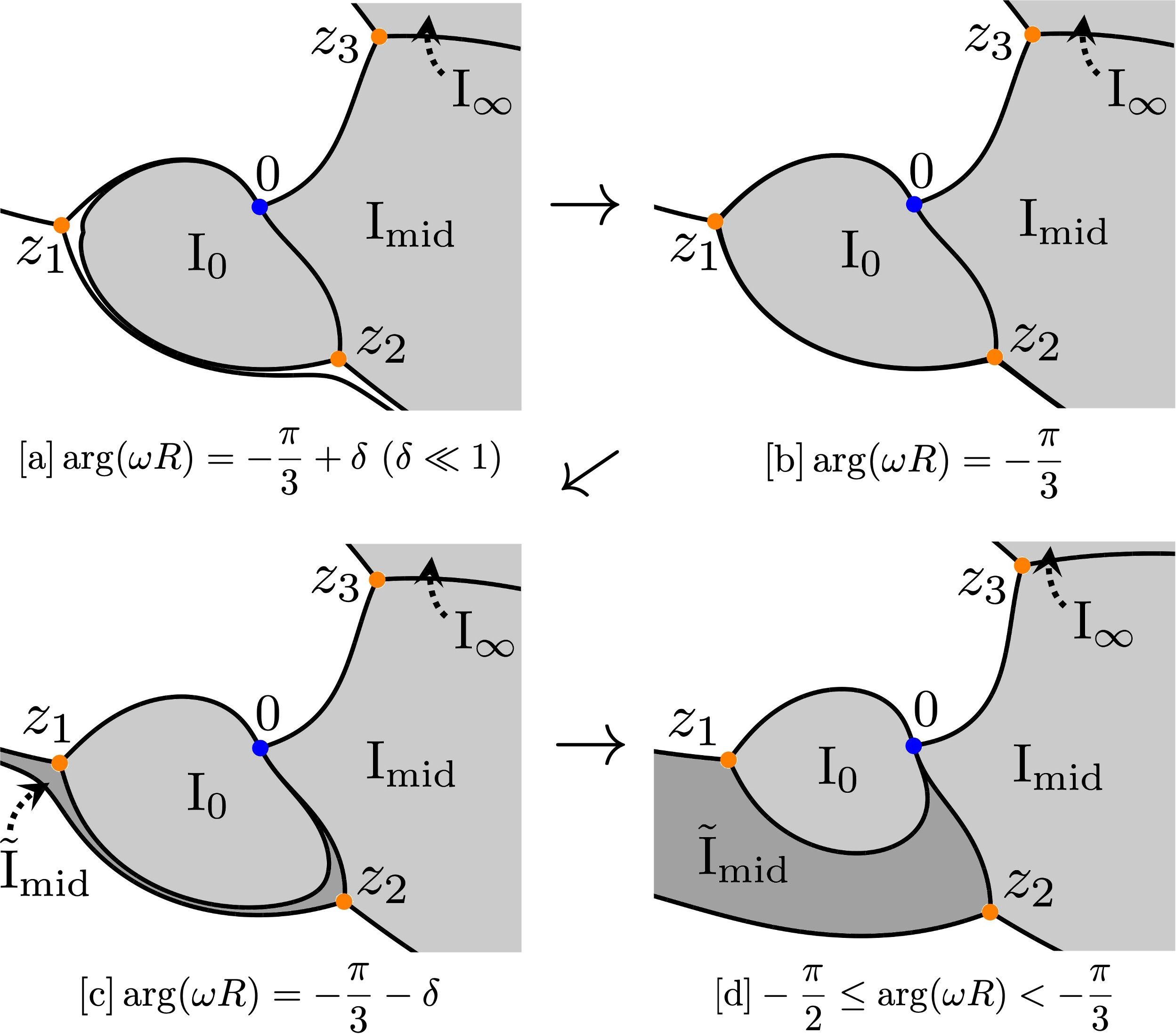}
    \caption{The saddle reduction of the Stokes graph.}
    \label{fig:tranSG}
  \end{center}
\end{figure}

Fig.\ref{fig:tranSG} shows the saddle reduction at $\arg(\omega R) = -\frac{\pi}{3}$ with $|\omega R/\lr{l + \frac{3}{2}}| = |\frac{2}{3} - i\frac{2}{30}|$. In the graph $\mathrm{[a]}$, the state of the adjacent of the Stokes regions does not change from Fig.\ref{fig:SGM5} (the Stokes region $\mathrm{I}_{\infty}$ is adjacent to $\mathrm{I}_{\mathrm{mid}}$, and $\mathrm{I}_{\mathrm{mid}}$ is adjacent to $\mathrm{I}_{0}$). In the graph $\mathrm{[b]}$, two Stokes lines emanating from $z_1$ and $z_2$ overlap and become one Stokes line connecting $z_1$ and $z_2$. Continuing to decrease $\arg(\omega R)$, we obtain the graph $\mathrm{[c]}$ and $\mathrm{[d]}$. In these cases, the Stokes line connecting $z_1$ and $z_2$ in $\mathrm{[b]}$ separates again but their positions invert. The new Stokes region $\tilde{\mathrm{I}}_{\mathrm{mid}}$ then appears between $\mathrm{I}_{\mathrm{mid}}$ and $\mathrm{I}_{0}$. For $\arg(\eta) = 0$ and $-\frac{\pi}{2} \leq \arg(\omega R) < -\frac{\pi}{3}$, the Stokes regions are adjacent each other in the same way as $\mathrm{[c]}$ and $\mathrm{[d]}$ in Fig.\ref{fig:tranSG}.

\subsection{Connection problems of WKB solutions}
\label{subsec:CP}

We solve the connection problem of the Borel resummed WKB solutions for $\arg(\eta) = 0$ from $z = \infty$ to the origin based on the Stokes graphs.

\subsubsection*{Connection problem for $-\frac{\pi}{3} < \arg(\omega R) < 0$}

\begin{figure}[t]
  \begin{center}
    \includegraphics[clip, width=10.0cm]{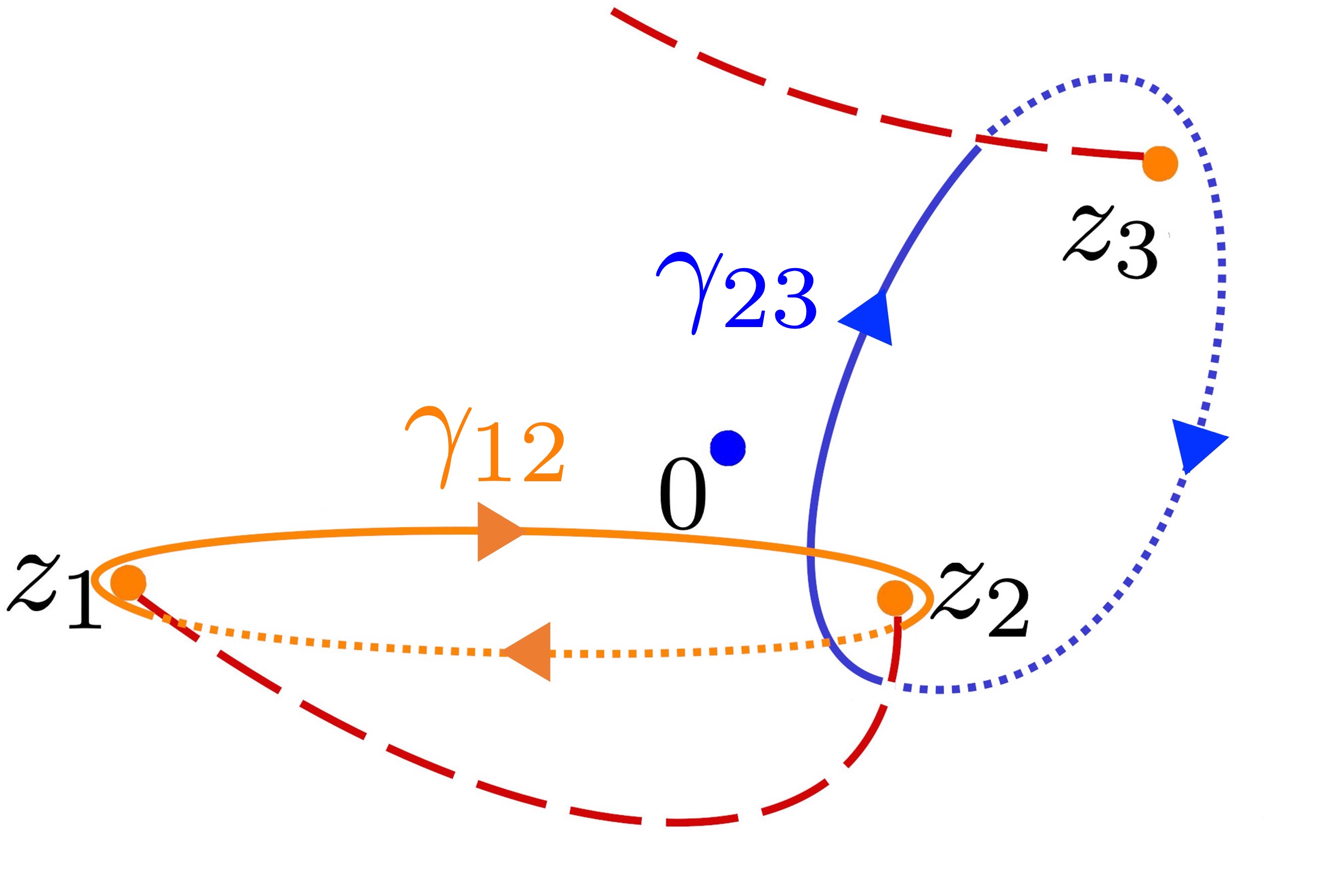}
    \caption{The one-cycles on $\Sigma$. The red dashed lines are the branch cuts. The solid segment in each cycle lies on the sheet of $\Sigma$ on where $P_{\mathrm{even}}(z)dz$ has the positive sign. The dashed segment in each cycle lies on the sheet of $\Sigma$ on where $P_{\mathrm{even}}(z)dz$ has the negative sign. The arrows on the cycles indicate the orientation.}
    \label{fig:cycles}
  \end{center}
\end{figure}

For $-\frac{\pi}{3} < \arg(\omega R) < 0$, we obtain the Stokes graph of the type Fig.\ref{fig:SGM5}. The connection problem from infinity to the origin is solved by computing the connection of the solutions from the Stokes region $\mathrm{I}_{\mathrm{\infty}}$ to $\mathrm{I}_{\mathrm{mid}}$ and $\mathrm{I}_{\mathrm{mid}}$ to $\mathrm{I}_{0}$. Between $\mathrm{I}_{\infty}$ and $\mathrm{I}_{\mathrm{mid}}$, there is a Stokes line emanating from $z_3$ and oriented to the positive direction. Then it is shown that the Borel resummed WKB solutions in the regions $\mathrm{I}_{\infty}$ and $\mathrm{I}_{\mathrm{mid}}$ are connected by the following connection formula \cite{DP, cr1, 2008.00379},
\beq
\label{eq:infmid}
\left( \begin{array}{r}
\mathcal{B}\lrk{\psi_{z_3, \mathrm{I_{\infty}}}^{+}}(z) \\ \mathcal{B}\lrk{\psi_{z_3, \mathrm{I_{\infty}}}^{-}}(z)
\end{array} \right) =  
\left( \begin{array}{rr}
1 & -i \\ 0 & 1
\end{array} \right) \left( \begin{array}{r}
\mathcal{B}\lrk{\psi_{z_3, \mathrm{I}_{\mathrm{mid}}}^{+}}(z) \\ \mathcal{B}\lrk{\psi_{z_3, \mathrm{I}_{\mathrm{mid}}}^{-}}(z)
\end{array} \right),
\eeq
where we have denoted the solutions in $\mathrm{I}_{\infty}$ as $\psi_{z_3, \mathrm{I_{\infty}}}^{\pm}(z)$ ($\pm$ in the superscript is the sign of $\pm P_{\mathrm{even}}(z)dz$ and $z_3$ in the subscript is the normalization point), and the solutions in $\mathrm{I}_{\mathrm{mid}}$ in the same way. Next, between $\mathrm{I}_{\mathrm{mid}}$ and $\mathrm{I}_{0}$, there is a Stokes line emanating from $z_2$ and oriented to the negative direction. In order to apply the connection formula, we need to change the normalization of the solutions,
\beq
\label{eq:midnor}
\left( \begin{array}{r}
\mathcal{B}\lrk{\psi_{z_3, \mathrm{I}_{\mathrm{mid}}}^{+}}(z) \\ \mathcal{B}\lrk{\psi_{z_3, \mathrm{I}_{\mathrm{mid}}}^{-}}(z)
\end{array} \right) =  
\left( \begin{array}{cc}
e^{-\eta\mathcal{B}\lrk{\int_{z_2}^{z_3}P_{\mathrm{even}}(z)dz}} & 0 \\ 0 & e^{\eta\mathcal{B}\lrk{\int_{z_2}^{z_3}P_{\mathrm{even}}(z)dz}}
\end{array} \right) \left( \begin{array}{r}
\mathcal{B}\lrk{\psi_{z_2, \mathrm{I}_{\mathrm{mid}}}^{+}}(z) \\ \mathcal{B}\lrk{\psi_{z_2, \mathrm{I}_{\mathrm{mid}}}^{-}}(z)
\end{array} \right),
\eeq
where we have used the property that the Borel resummation commutes with addition and multiplication  (\ref{eq:Bsumalge}). After changing the normalization, the solutions in $\mathrm{I}_{\mathrm{mid}}$ and $\mathrm{I}_{0}$ are connected as follows \cite{DP, cr1, 2008.00379},
\beq
\label{eq:mid0}
\left( \begin{array}{r}
\mathcal{B}\lrk{\psi_{z_2, \mathrm{I}_{\mathrm{mid}}}^{+}}(z) \\ \mathcal{B}\lrk{\psi_{z_2, \mathrm{I}_{\mathrm{mid}}}^{-}}(z)
\end{array} \right) =  
\left( \begin{array}{rr}
1 & 0 \\ i & 1
\end{array} \right) \left( \begin{array}{r}
\mathcal{B}\lrk{\psi_{z_2, \mathrm{I}_{0}}^{+}}(z) \\ \mathcal{B}\lrk{\psi_{z_2, \mathrm{I}_{0}}^{-}}(z)
\end{array} \right).
\eeq
By multiplying (\ref{eq:infmid})$\sim$(\ref{eq:mid0}) and finally restoring the normalization of $\psi_{z_2, \mathrm{I}_{0}}^{\pm}$ in the r.h.s. of (\ref{eq:mid0}) from $z_2$ to $z_3$, we obtain the solution of the connection problem for $-\frac{\pi}{3} < \arg(\omega R) < 0$,
\beq
\label{eq:intCF}
\left( \begin{array}{r}
\mathcal{B}\lrk{\psi_{z_3, \mathrm{I_{\infty}}}^{+}}(z) \\ \mathcal{B}\lrk{\psi_{z_3, \mathrm{I_{\infty}}}^{-}}(z)
\end{array} \right) =  
\left( \begin{array}{rr}
1 + e^{\eta\mathcal{B}\lrk{2\int_{z_2}^{z_3}P_{\mathrm{even}}(z)dz}} & -i \\ ie^{\eta\mathcal{B}\lrk{2\int_{z_2}^{z_3}P_{\mathrm{even}}(z)dz}} & 1
\end{array} \right) \left( \begin{array}{r}
\mathcal{B}\lrk{\psi_{z_3, \mathrm{I_{0}}}^{+}}(z) \\ \mathcal{B}\lrk{\psi_{z_3, \mathrm{I_{0}}}^{-}}(z)
\end{array} \right).
\eeq
The integral $2\int_{z_2}^{z_3}P_{\mathrm{even}}(z)dz$ is equivalent to the WKB period for the one-cycle $\gamma_{23}$ in Fig.\ref{fig:cycles}. Therefore (\ref{eq:intCF}) can be expressed in terms of the WKB period,
\beq
\label{eq:periodCF}
\left( \begin{array}{r}
\mathcal{B}\lrk{\psi_{z_3, \mathrm{I_{\infty}}}^{+}}(z) \\ \mathcal{B}\lrk{\psi_{z_3, \mathrm{I_{\infty}}}^{-}}(z)
\end{array} \right) =  
\left( \begin{array}{rr}
1 + e^{\eta\mathcal{B}\lrk{\Pi_{\gamma_{23}}}} & -i \\ ie^{\eta\mathcal{B}\lrk{\Pi_{\gamma_{23}}}} & 1
\end{array} \right) \left( \begin{array}{r}
\mathcal{B}\lrk{\psi_{z_3, \mathrm{I_{0}}}^{+}}(z) \\ \mathcal{B}\lrk{\psi_{z_3, \mathrm{I_{0}}}^{-}}(z)
\end{array} \right).
\eeq

\subsubsection*{Connection problem for $-\frac{\pi}{2} \leq \arg(\omega R) < -\frac{\pi}{3}$}
\begin{figure}[t]
  \begin{center}
    \includegraphics[clip, width=10.0cm]{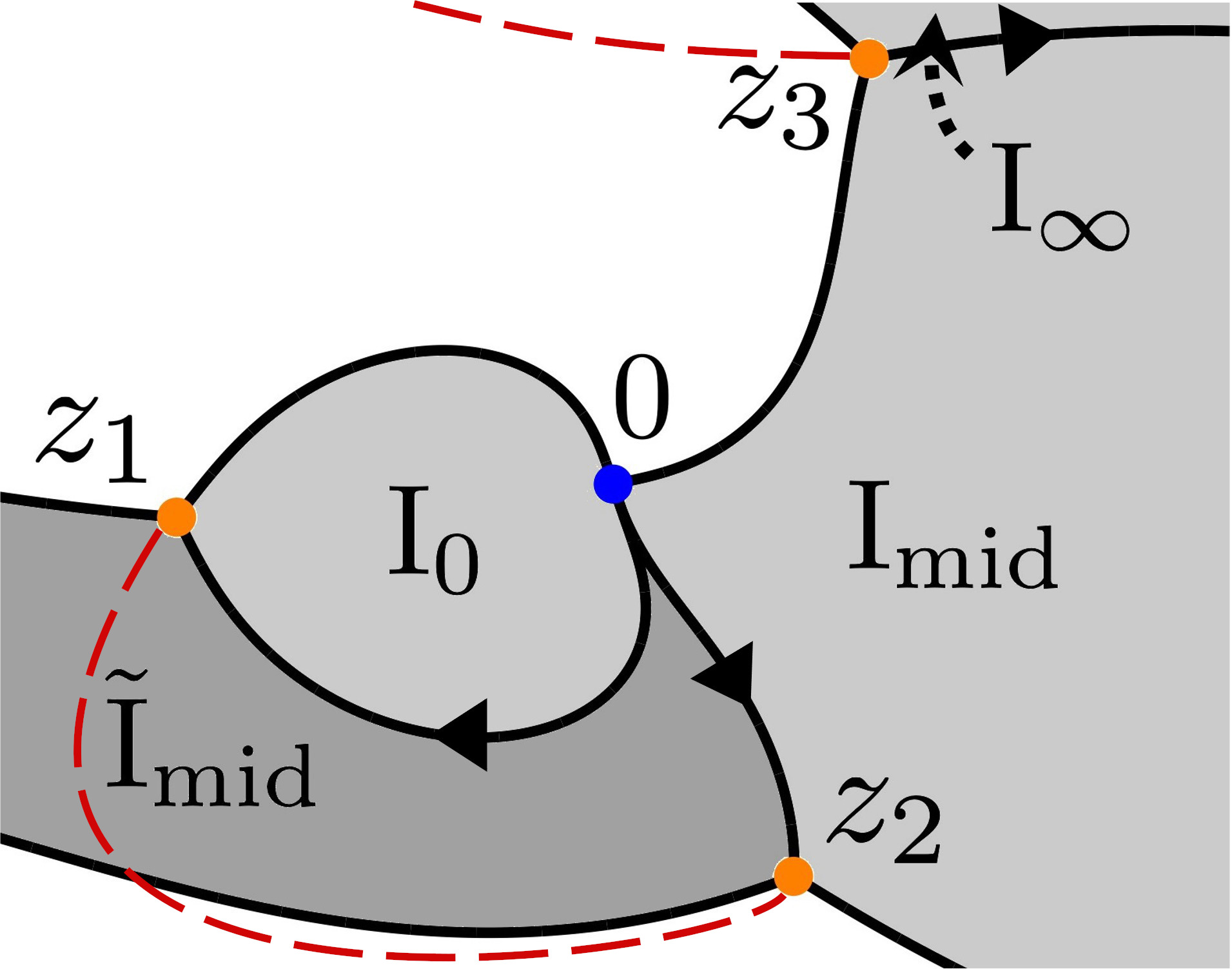}
    \caption{The Stokes graph for $\arg(\eta) = 0$ and $-\frac{\pi}{2} \leq \arg(\omega R) < -\frac{\pi}{3}$.}
    \label{fig:SGM5flip}
  \end{center}
\end{figure}

For $-\frac{\pi}{2} \leq \arg(\omega R) < -\frac{\pi}{3}$, we obtain the Stokes graph of the type $\mathrm{[c]}$ and $\mathrm{[d]}$ in Fig.\ref{fig:tranSG}. The orientation of the Stokes lines and the branch cuts on $\Sigma$ is drawn in Fig.\ref{fig:SGM5flip}.

The connection problem from infinity to the origin is solved by computing the connection of the solutions from $\mathrm{I}_{\mathrm{\infty}}$ to $\mathrm{I}_{\mathrm{mid}}$, $\mathrm{I}_{\mathrm{mid}}$ to $\tilde{\mathrm{I}}_{\mathrm{mid}}$, and $\tilde{\mathrm{I}}_{\mathrm{mid}}$ to $\mathrm{I}_{0}$ successively. The connection formula from $\mathrm{I}_{\mathrm{\infty}}$ to $\mathrm{I}_{\mathrm{mid}}$ is same as (\ref{eq:infmid}). Changing the normalization of the solutions from $z_3$ to $z_2$ has also already given by (\ref{eq:midnor}). The connection formula from $\mathrm{I}_{\mathrm{mid}}$ to $\tilde{\mathrm{I}}_{\mathrm{mid}}$ is given by (\ref{eq:mid0}) with replacing $\mathrm{I}_{0}$ in the r.h.s. to $\tilde{\mathrm{I}}_{\mathrm{mid}}$,
\beq
\label{eq:midtil}
\left( \begin{array}{r}
\mathcal{B}\lrk{\psi_{z_2, \mathrm{I}_{\mathrm{mid}}}^{+}}(z) \\ \mathcal{B}\lrk{\psi_{z_2, \mathrm{I}_{\mathrm{mid}}}^{-}}(z)
\end{array} \right) =  
\left( \begin{array}{rr}
1 & 0 \\ i & 1
\end{array} \right) \left( \begin{array}{r}
\mathcal{B}\lrk{\psi_{z_2, \tilde{\mathrm{I}}_{\mathrm{mid}}}^{+}}(z) \\ \mathcal{B}\lrk{\psi_{z_2, \tilde{\mathrm{I}}_{\mathrm{mid}}}^{-}}(z)
\end{array} \right).
\eeq
To compute the connection from $\tilde{\mathrm{I}}_{\mathrm{mid}}$ to $\mathrm{I}_{0}$, we need to change the normalization of the solutions from $z_2$ to $z_1$,
\beq
\label{eq:tilnor}
\left( \begin{array}{r}
\mathcal{B}\lrk{\psi_{z_2, \tilde{\mathrm{I}}_{\mathrm{mid}}}^{+}}(z) \\ \mathcal{B}\lrk{\psi_{z_2, \tilde{\mathrm{I}}_{\mathrm{mid}}}^{-}}(z)
\end{array} \right) =  
\left( \begin{array}{cc}
e^{-\eta\mathcal{B}\lrk{\int_{z_1}^{z_2}P_{\mathrm{even}}(z)dz}} & 0 \\ 0 & e^{\eta\mathcal{B}\lrk{\int_{z_1}^{z_2}P_{\mathrm{even}}(z)dz}}
\end{array} \right) \left( \begin{array}{r}
\mathcal{B}\lrk{\psi_{z_1, \tilde{\mathrm{I}}_{\mathrm{mid}}}^{+}}(z) \\ \mathcal{B}\lrk{\psi_{z_1, \tilde{\mathrm{I}}_{\mathrm{mid}}}^{-}}(z)
\end{array} \right).
\eeq
Between $\tilde{\mathrm{I}}_{\mathrm{mid}}$ and $\mathrm{I}_{0}$, there is a Stokes line emanating from $z_1$ and oriented to the negative direction. Then the solutions are connected by the following formula,
\beq
\label{eq:til0}
\left( \begin{array}{r}
\mathcal{B}\lrk{\psi_{z_1, \tilde{\mathrm{I}}_{\mathrm{mid}}}^{+}}(z) \\ \mathcal{B}\lrk{\psi_{z_1, \tilde{\mathrm{I}}_{\mathrm{mid}}}^{-}}(z)
\end{array} \right) =  
\left( \begin{array}{rr}
1 & 0 \\ i & 1
\end{array} \right) \left( \begin{array}{r}
\mathcal{B}\lrk{\psi_{z_1, \mathrm{I}_{0}}^{+}}(z) \\ \mathcal{B}\lrk{\psi_{z_1, \mathrm{I}_{0}}^{-}}(z)
\end{array} \right).
\eeq
By multiplying (\ref{eq:infmid}), (\ref{eq:midnor}), (\ref{eq:midtil})$\sim$(\ref{eq:til0}) and finally restoring the normalization of $\psi_{z_1, \mathrm{I}_{0}}^{\pm}$ in the r.h.s. of (\ref{eq:til0}) from $z_1$ to $z_3$, we obtain the solution of the connection problem for $-\frac{\pi}{2} \leq \arg(\omega R) < -\frac{\pi}{3}$,
\beq
\label{eq:intCF2}
\left( \begin{array}{r}
\mathcal{B}\lrk{\psi_{z_3, \mathrm{I_{\infty}}}^{+}}(z) \\ \mathcal{B}\lrk{\psi_{z_3, \mathrm{I_{\infty}}}^{-}}(z)
\end{array} \right) =  
\left( \begin{array}{rr}
1 + \left(1 + e^{\eta\mathcal{B}\lrk{2\int_{z_1}^{z_2}P_{\mathrm{even}}(z)dz}}\right)e^{\eta\mathcal{B}\lrk{2\int_{z_2}^{z_3}P_{\mathrm{even}}(z)dz}} & -i \\ i\left(1 + e^{\eta\mathcal{B}\lrk{2\int_{z_1}^{z_2}P_{\mathrm{even}}(z)dz}}\right)e^{\eta\mathcal{B}\lrk{2\int_{z_2}^{z_3}P_{\mathrm{even}}(z)dz}} & 1
\end{array} \right) \left( \begin{array}{r}
\mathcal{B}\lrk{\psi_{z_3, \mathrm{I_{0}}}^{+}}(z) \\ \mathcal{B}\lrk{\psi_{z_3, \mathrm{I_{0}}}^{-}}(z)
\end{array} \right).
\eeq
The integrals $2\int_{z_2}^{z_3}P_{\mathrm{even}}(z)dz$ and $2\int_{z_1}^{z_2}P_{\mathrm{even}}(z)dz$ are equivalent to the WKB periods for the one-cycles $\gamma_{23}$ and $\gamma_{12}$ in Fig.\ref{fig:cycles}, respectively. Therefore (\ref{eq:intCF2}) can be expressed in terms of the WKB periods,
\beq
\label{eq:periodCF2}
\left( \begin{array}{r}
\mathcal{B}\lrk{\psi_{z_3, \mathrm{I_{\infty}}}^{+}}(z) \\ \mathcal{B}\lrk{\psi_{z_3, \mathrm{I_{\infty}}}^{-}}(z)
\end{array} \right) =  
\left( \begin{array}{rr}
1 + \left(1 + e^{\eta\mathcal{B}\lrk{\Pi_{\gamma_{12}}}}\right)e^{\eta\mathcal{B}\lrk{\Pi_{\gamma_{23}}}} & -i \\ i\left(1 + e^{\eta\mathcal{B}\lrk{\Pi_{\gamma_{12}}}}\right)e^{\eta\mathcal{B}\lrk{\Pi_{\gamma_{23}}}} & 1
\end{array} \right) \left( \begin{array}{r}
\mathcal{B}\lrk{\psi_{z_3, \mathrm{I_{0}}}^{+}}(z) \\ \mathcal{B}\lrk{\psi_{z_3, \mathrm{I_{0}}}^{-}}(z)
\end{array} \right).
\eeq

\subsection{QNMs conditions for extremal M5-branes}
\label{subsec:QNMcond}
We impose the boundary conditions (i) only outgoing wave exists at $z \rightarrow +\infty$ and (ii) only ingoing wave exists at $z \rightarrow 0$ on the Borel resummed WKB solutions. The outgoing wave at $z \rightarrow +\infty$ is $\mathcal{B}\lrk{\psi_{z_3, \mathrm{I_{\infty}}}^{+}}(z)$. According to the connection formulae (\ref{eq:periodCF}) and (\ref{eq:periodCF2}), $\mathcal{B}\lrk{\psi_{z_3, \mathrm{I_{\infty}}}^{+}}(z)$ is connected to the origin as follows,
\beq
\label{eq:BCconnect}
\mathcal{B}\lrk{\psi_{z_3, \mathrm{I_{\infty}}}^{+}}(z) = 
\left\{
\begin{array}{l}
\lr{1 + e^{\eta\mathcal{B}\lrk{\Pi_{\gamma_{23}}}}}\mathcal{B}\lrk{\psi_{z_3, \mathrm{I_{0}}}^{+}}(z) -i\mathcal{B}\lrk{\psi_{z_3, \mathrm{I_{0}}}^{-}}(z)\ \ \left(-\frac{\pi}{3} < \arg(\omega R) < 0\right), \\
\lr{1 + \left(1 + e^{\eta\mathcal{B}\lrk{\Pi_{\gamma_{12}}}}\right)e^{\eta\mathcal{B}\lrk{\Pi_{\gamma_{23}}}}}\mathcal{B}\lrk{\psi_{z_3, \mathrm{I_{0}}}^{+}}(z) -i\mathcal{B}\lrk{\psi_{z_3, \mathrm{I_{0}}}^{-}}(z)\\
\ \ \ \ \ \ \ \ \ \ \ \ \ \ \ \ \ \ \ \ \ \ \ \ \ \ \ \ \ \ \ \ \ \ \ \ \ \ \ \ \ \ \ \ \ \ \ \ \ \ \ \ \ \ \ \ \ \ \ \left(-\frac{\pi}{2} \leq \arg(\omega R) < -\frac{\pi}{3}\right).
\end{array}
\right.
\eeq
The ingoing wave at $z \rightarrow 0$ is $\mathcal{B}\lrk{\psi_{z_3, \mathrm{I_{0}}}^{-}}(z)$. Therefore to satisfy the boundary conditions, the coefficients of $\mathcal{B}\lrk{\psi_{z_3, \mathrm{I_{0}}}^{+}}(z)$ in (\ref{eq:BCconnect}) need to be zero,
\beq
\label{eq:BCcoef}
\left\{
\begin{array}{l}
1 + e^{\eta\mathcal{B}\lrk{\Pi_{\gamma_{23}}}} = 0\ \ \ \ \ \ \ \ \ \ \ \ \ \ \ \ \ \ \ \ \ \ \left(-\frac{\pi}{3} < \arg(\omega R) < 0\right), \\
1 + \left(1 + e^{\eta\mathcal{B}\lrk{\Pi_{\gamma_{12}}}}\right)e^{\eta\mathcal{B}\lrk{\Pi_{\gamma_{23}}}} = 0\ \ \left(-\frac{\pi}{2} \leq \arg(\omega R) < -\frac{\pi}{3}\right).
\end{array}
\right.
\eeq
Taking the logarithm of (\ref{eq:BCcoef}) and substituting $\eta = 1$, we obtain the following conditions,
\beq
\label{eq:exactQNM}
\left\{
\begin{array}{l}
\mathcal{B}\lrk{\Pi_{\gamma_{23}}} = 2\pi i\left(n + \frac{1}{2}\right)\ \ \ \ \ \ \ \ \ \ \ \ \ \ \ \ \ \ \ \ \ \ \ \ \ \ \ \ \left(-\frac{\pi}{3} < \arg(\omega R) < 0\right), \\
\mathcal{B}\lrk{\Pi_{\gamma_{23}}} + \log\left(1 + e^{\mathcal{B}\lrk{\Pi_{\gamma_{12}}}}\right) = 2\pi i\left(n + \frac{1}{2}\right)\ \ \left(-\frac{\pi}{2} \leq \arg(\omega R) < -\frac{\pi}{3}\right),
\end{array}
\right.
\eeq
where $n \in \mathbb{Z}_{\geq 0}$. The conditions (\ref{eq:exactQNM}) are satisfied by discrete sets of $\omega$, which are the quasi-normal modes for the massless scalar perturbation in the extremal M5-brane metric. Without considering convergence of the asymptotic series, the asymptotic expansion of (\ref{eq:exactQNM}) provides the following conditions,
\beq
\label{eq:asymQNM}
\left\{
\begin{array}{l}
\Pi_{\gamma_{23}} = 2\pi i\left(n + \frac{1}{2}\right)\ \ \ \ \ \ \ \ \ \ \ \ \ \ \ \ \ \ \ \ \ \ \ \ \left(-\frac{\pi}{3} < \arg(\omega R) < 0\right), \\
\Pi_{\gamma_{23}} + \log\left(1 + e^{\Pi_{\gamma_{12}}}\right) = 2\pi i\left(n + \frac{1}{2}\right)\ \ \left(-\frac{\pi}{2} \leq \arg(\omega R) < -\frac{\pi}{3}\right).
\end{array}
\right.
\eeq
The first condition of (\ref{eq:asymQNM}) is the all-order extension of the Bohr-Sommerfeld condition (\ref{eq:largeO}).

\vspace{0.2in}
\par
\begin{figure}[t]
  \begin{center}
    \includegraphics[clip, width=12.9cm]{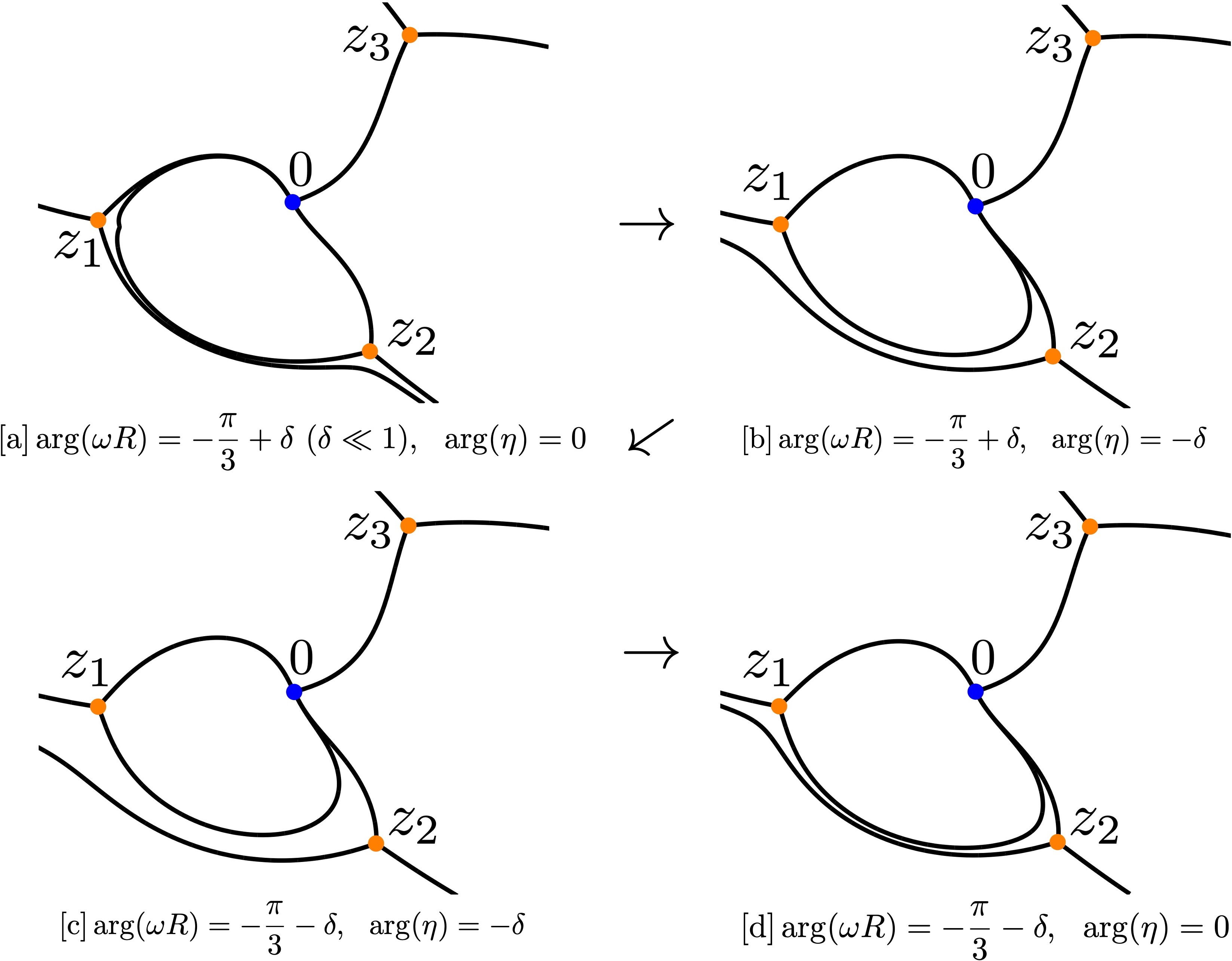}
    \caption{The transition of the Stokes graph with rotating $\arg(\omega R)$ and $\arg(\eta)$. The graph $\mathrm{[a]}$, $\mathrm{[d]}$ are same as $\mathrm{[a]}$, $\mathrm{[d]}$ in Fig.\ref{fig:tranSG}, respectively.}
    \label{fig:SGDP}
  \end{center}
\end{figure}

We present some comments for (\ref{eq:BCcoef})$\sim$(\ref{eq:asymQNM}) from the point of view of the resurgence. First, the two QNMs conditions are continuous. Let us see the transition of the Stokes graph with rotating $\arg(\omega R)$ and $\arg(\eta)$ (Fig.\ref{fig:SGDP}). From the graph $\mathrm{[a]}$ to $\mathrm{[b]}$ in Fig.\ref{fig:SGDP}, $\arg(\eta)$ is rotated from $\arg(\eta) = 0$ to $\arg(\eta) = -\delta$ while $\arg(\omega R) = -\pi/3 + \delta$ is fixed. At an argument $\arg(\eta) = -\varphi$ $(-\delta < -\varphi < 0)$, the Stokes line connecting $z_1$ and $z_2$ appears and the saddle reduction happens. Then the Borel transformation of the WKB period $\Pi_{\gamma_{23}}$, whose one-cycle $\gamma_{23}$ intersects with the Stokes line connecting $z_1$ and $z_2$, has a singularity on the complex $\xi$-plane in the direction $\arg(\xi) = +\varphi$. If we rotate $\arg(\eta)$ as crossing $\arg(\eta) = -\varphi$, the integration contour of the Borel resummation for $\Pi_{\gamma_{23}}$ collides with the singularity on the complex $\xi$-plane. To ensure the analyticity in $\eta$ of the Borel resummation, we must take the contribution from the integration contour encircling the singularity into account. Then it is shown that the WKB period is connected as follows (theorem 2.5.1 of \cite{DP}, and theorem 3.4 of \cite{cr1}),
\beq
\label{eq:DP}
\lim_{\arg(\eta) \rightarrow -\varphi +0} e^{\eta\mathcal{B}\lrk{\Pi_{\gamma_{23}}}} = \lim_{\arg(\eta) \rightarrow -\varphi -0}\left(1 + e^{\eta\mathcal{B}\lrk{\Pi_{\gamma_{12}}}}\right)e^{\eta\mathcal{B}\lrk{\Pi_{\gamma_{23}}}}.
\eeq
This is called Delabaere-Pham formula. The term $e^{\eta\mathcal{B}\lrk{\Pi_{\gamma_{12}}}}e^{\eta\mathcal{B}\lrk{\Pi_{\gamma_{23}}}}$ in r.h.s. of (\ref{eq:DP}) is the contribution from the singularity of the Borel transformation of $\Pi_{\gamma_{23}}$.

\par
From the graph $\mathrm{[b]}$ to $\mathrm{[c]}$ in Fig.\ref{fig:SGDP}, $\arg(\omega R)$ is rotated from $\arg(\omega R) = -\pi/3 + \delta$ to $\arg(\omega R) = -\pi/3 -\delta$ while $\arg(\eta) = -\delta$ is fixed.  Finally from the graph $\mathrm{[c]}$ to $\mathrm{[d]}$, $\arg(\eta)$ is rotated from $\arg(\eta) = -\delta$ to $\arg(\eta) = 0$ while $\arg(\omega R) = -\pi/3 -\delta$ is fixed. During $\mathrm{[b]}$ to $\mathrm{[d]}$, the saddle reduction does not happen and therefore the r.h.s. of (\ref{eq:DP}) does not change discontinuously. In summary, Fig.\ref{fig:SGDP} indicates $e^{\eta\mathcal{B}\lrk{\Pi_{\gamma_{23}}}}$ for $-\frac{\pi}{3} < \arg(\omega R) < 0$ and $\left(1 + e^{\eta\mathcal{B}\lrk{\Pi_{\gamma_{12}}}}\right)e^{\eta\mathcal{B}\lrk{\Pi_{\gamma_{23}}}}$ for $-\frac{\pi}{2} \leq \arg(\omega R) < -\frac{\pi}{3}$ with $\arg(\eta) = 0$ are continuous. Therefore the two conditions in (\ref{eq:BCcoef}) are also continuous.
\par
Next comment is about the non-perturbative property of the QNMs. Calculating the QNMs from the conditions (\ref{eq:BCcoef})$\sim$(\ref{eq:asymQNM}), one finds the term $e^{\eta\mathcal{B}\lrk{\Pi_{\gamma_{12}}}}$ in the conditions for $-\frac{\pi}{2} \leq \arg(\omega R) < -\frac{\pi}{3}$ is exponentially small. We can also find this type of exponentially small terms in the study of the energy quantization conditions for resurgent quantum mechanics (e.g. \cite{Esemi, quant-ph/0501136, quant-ph/0501137, 1501.05671, 2103.06586}). In these cases, the exponentially small terms arise from the tunnel effect and produce the instanton contributions to the asymptotic expansion of the energy eigenvalues. The conditions (\ref{eq:BCcoef})$\sim$(\ref{eq:asymQNM}) imply that the QNMs in the region $-\frac{\pi}{2} \leq \arg(\omega R) < -\frac{\pi}{3}$ have such an instanton contribution. 
\par
Following the argument for the non-perturbative property, the continuation for the QNMs conditions (\ref{eq:DP}) can be seen as that the discontinuity of the perturbative part is leading the non-perturbative part of the QNMs.

\section{Computations of QNMs}
\label{sec:CompQNM}
In this section, we will numerically and analytically compute the QNMs by using the conditions (\ref{eq:BCcoef})$\sim$(\ref{eq:asymQNM}). We first compute the WKB periods. The leading order contributions of the WKB periods are given by two times (\ref{eq:13int})$\sim$(\ref{eq:12int}).  To compute the higher order contributions, it is useful to transform the equation (\ref{eq:Schro}). Under the following transformations, 
\beq
\label{eq:SWGdict}
\begin{split}
&\ \ \ \ \ \ \ \ \ \ \ \eta = \frac{1}{\hbar}, \ \ \left(l + \frac{3}{2}\right)^2 = u, \ \ \omega R = i\frac{\Lambda_1}{2}, \\
&z = (\sqrt{\Lambda_1}/2)e^{ix}, \ \ \psi(z(x)) = (\sqrt{\Lambda_1}/2)^{(1/2)}e^{i\frac{x}{2}}\Psi(x),
\end{split}
\eeq
the equation (\ref{eq:Schro}) becomes as follows,
\beq
\label{eq:QSWC}
\lrk{\hbar^2\frac{d^2}{dx^2} + u + \frac{1}{16}\Lambda_1^3e^{2ix}+\frac{1}{2}\Lambda_1^{\frac{3}{2}}e^{-ix}}\Psi(x) = 0.
\eeq
(\ref{eq:QSWC}) is the quantum Seiberg-Witten curve for 4-dimensional $\mathcal{N} = 2$ SU(2) supersymmetric QCD with a massless fundamental matter \cite{1705.09120}. In the context of the gauge theory, $u$ is the Coulomb moduli parameter, $\Lambda_1$ is the dynamically generated scale, and $\hbar$ is the deformation parameter in the Nekrasov-Shatashvili limit of the $\Omega$-background. (\ref{eq:SWGdict}) is a new example of the Seiberg-Witten/gravity correspondence \cite{2006.06111, 2105.04245, 2109.09804}. Under the transformation (\ref{eq:SWGdict}), it is shown that the WKB periods are invariant (Proposition 2.7 (b) in \cite{cr1}). Therefore computing the WKB periods for (\ref{eq:Schro}) is equivalent to computing the WKB periods for the quantum Seiberg-Witten curve (\ref{eq:QSWC}). The higher-order contributions of the WKB periods for the quantum Seiberg-Witten curve can be computed by applying differential operators with respect to $u$ to the leading order contributions \cite{1705.09120},
\beq
 \Pi_{\gamma}^{(2n)} = \mathcal{O}_{2n}\Pi_{\gamma}^{(0)}.
\eeq
Up to 4th-order, the differential operators are given as follows,
\beq
\label{eq:difop}
\begin{split}
 &\mathcal{O}_{2} = \frac{1}{12}\left(2u\frac{\partial^2}{\partial u^2} + \frac{\partial}{\partial u}\right),\\
 &\mathcal{O}_{4} = \frac{1}{1440}\left(28u^2\frac{\partial^4}{\partial u^4} + 124u\frac{\partial^3}{\partial u^3} + 81\frac{\partial^2}{\partial u^2}\right).
\end{split}
\eeq
\par

\begin{figure}[t]
  \begin{center}
    \includegraphics[clip, width=11.5cm]{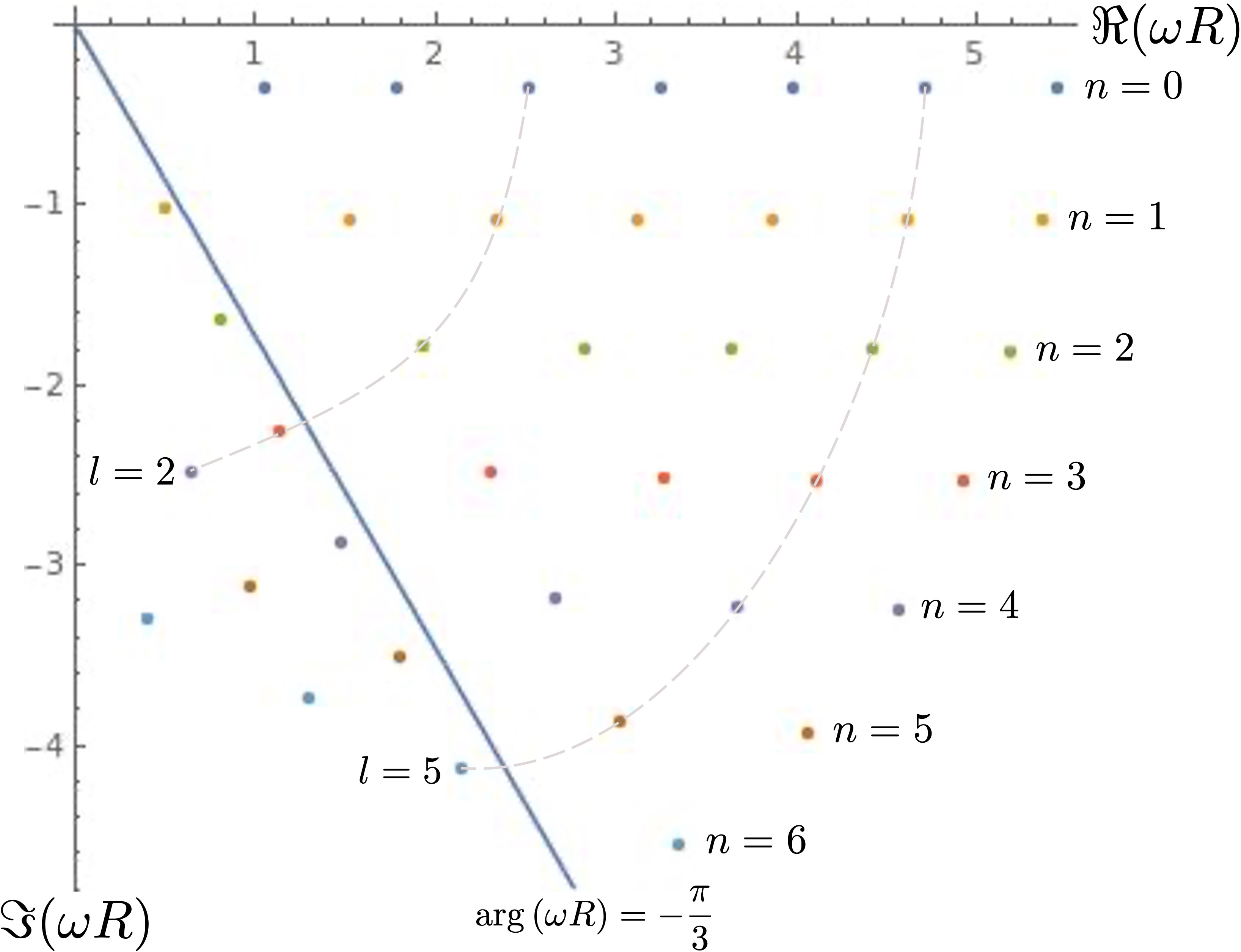}
    \caption{The numerical values of the QNMs calculated from the exact conditions (\ref{eq:exactQNM}). The upper right points from the line at $\arg(\omega R) = -\frac{\pi}{3}$ are calculated by the condition for $-\frac{\pi}{3} < \arg(\omega R) < 0$ of (\ref{eq:exactQNM}), and the lower left points are calculated by the condition for $-\frac{\pi}{3} < \arg(\omega R) < 0$.}
    \label{fig:QNMs}
  \end{center}
\end{figure}

After computing the WKB periods, we take the Borel resummation of them. In this paper, we use the Borel-Pad\'{e} approximation,
\beq
 \mathcal{B}\lrk{\Pi_{\gamma}} \sim \int_{0}^{\infty}e^{-\xi}\lrk{N/N}d\xi,
\eeq
where $\lrk{N/N}$ is the diagonal Pad\'{e} approximation of order $N$ for the Borel transformation of the WKB periods. 
\par
Fig.\ref{fig:QNMs} shows the numerical values of the QNMs calculated from the exact conditions (\ref{eq:exactQNM}), where we have used up to 4th-order of the WKB periods and the diagonal Pad\'{e} approximation of order $2$. In Fig.\ref{fig:QNMs}, the QNMs for $l < n$ are located in $\arg(\omega R) < -\frac{\pi}{3}$ region. With fixed $l$, the imaginary part of the QNMs becomes small as $n$ increases. The polarization modes with $\bm{k} = 0$ of the scalar field then get to decay at $t \rightarrow +\infty$ quickly. We also compare the results of (\ref{eq:exactQNM}) with the large $l$ expression (\ref{eq:largeOappAF}) in Table.\ref{tab:BSvsEQC}. At $n = 1$ and $l = 0$, the argument of the numerical result of (\ref{eq:exactQNM}) is smaller than $-\frac{\pi}{3}$ while the argument of the result of (\ref{eq:largeOappAF}) is larger than $-\frac{\pi}{3}$. This result indicates that we need to use the second condition in (\ref{eq:exactQNM}) to calculate the QNMs in $\arg(\omega R) < -\frac{\pi}{3}$ region (or $l < n$ region).

\setlength\dashlinedash{.4pt}
\setlength\dashlinegap{1.3pt}
\begin{table}[t]
\begin{center}
\begin{tabular}{c|c:cc:cc:} \hline
$n$ & $l$ & (\ref{eq:exactQNM}) & $-\frac{\pi}{3} < \arg(\omega R)$ & (\ref{eq:largeOappAF}) & $-\frac{\pi}{3} < \arg(\omega R)$  \\ \hline\hline
0 & 0 & $1.057134 - 0.365624i$ & True & $1.028721 - 0.363707i$ & True \\
& 1 & $1.798272 - 0.364387i$ & True & $1.781797 - 0.363707i$ & Ture \\
& 2 & $2.531500 - 0.364053i$ & True & $2.519842 - 0.363707i$ & True \\ \hline\hline
1 & 0 & $0.506956 - 1.023014i$ & False & $1.028721 - 1.091123i$ & True \\
 & 1 & $1.535914 - 1.086901i$ & True & $1.781797 - 1.091123i$ & True \\
& 2 & $2.351632 - 1.089620i$ & True & $2.519842 - 1.091123i$ & True \\ \hline
\end{tabular}
\caption{Numerical results of the exact conditions (\ref{eq:exactQNM}) and the large $l$ expression (\ref{eq:largeOappAF}). The fourth column indicates which the numerical values of the third column satisfy $-\frac{\pi}{3} < \arg(\omega R)$ or not, and the sixth column indicates which the numerical values of the fifth column satisfy $-\frac{\pi}{3} < \arg(\omega R)$ or not.}
\label{tab:BSvsEQC}
\end{center}
\end{table}

\par
We would like to make a comment for the numerical calculation of the QNMs. To calculate the QNMs, we often use the Leaver's 3-term recurrence method \cite{Leaver}. But for the present case, we obtained a 7-term recurrence relation and we were not able to apply the Leaver's method. We also tried to use the matrix Leaver's method \cite{LeaverM, 2011.13859}, but we were not able to obtain the results with enough convergence. According to these reasons, it is thought that the exact WKB analysis is a convenient tool to study the QNMs for the M5-branes.
\vspace{0.2in}
\par
By using the Pad\'{e} approximation, one can numerically check the singularity of the Borel transformation of the WKB period. For example, at $\omega R = \exp\lr{-i\frac{\pi}{3} +i\frac{1}{100}}$ and $l = 0$, we find that the diagonal Pad\'{e} approximation of order $2$ for the Borel transformation of $\Pi_{\gamma_{23}}$ with 4th-order coefficients has a singularity at $\xi = 1.797240 + 0.320820i$.
\vspace{0.2in}
\par
Analytic expressions of the QNMs can also be obtained in parts. For $-\frac{\pi}{6} \leq \arg(\omega R) \leq 0$, the coefficients of the WKB period $\Pi_{\gamma_{23}}^{(n)}$ can be expanded at $\omega R = \frac{\sqrt[3]{2}\left(l + \frac{3}{2}\right)}{\sqrt{3}}$,
\beq
\label{eq:coefexp}
\begin{split}
 \Pi_{\gamma_{23}}^{(0)} &= -2^{2/3}\sqrt{3}\pi \left(\omega R - \frac{\sqrt[3]{2} \left(l + \frac{3}{2}\right)}{\sqrt{3}}\right)+\frac{5 \pi \left(\omega R-\frac{\sqrt[3]{2} \left(l + \frac{3}{2}\right)}{\sqrt{3}}\right)^2}{2^{5/3} \left(l + \frac{3}{2}\right)}+\cdots , \\
 \Pi_{\gamma_{23}}^{(2)} &= \frac{5 \pi }{72 \left(l + \frac{3}{2}\right)} - \frac{25 \pi \left(\omega R-\frac{\sqrt[3]{2} \left(l + \frac{3}{2}\right)}{\sqrt{3}}\right)}{288 \left(\sqrt[3]{2} \sqrt{3} \left(l + \frac{3}{2}\right)^2\right)} - \frac{325 \pi \left(\omega R-\frac{\sqrt[3]{2} \left(l + \frac{3}{2}\right)}{\sqrt{3}}\right)^2}{10368 \left(2^{2/3} \left(l + \frac{3}{2}\right)^3\right)}+\cdots ,  \\
 \Pi_{\gamma_{23}}^{(4)} &= -\frac{2471 \pi }{746496 \left(l + \frac{3}{2}\right)^3} + \frac{33565 \pi  \left(\omega R-\frac{\sqrt[3]{2} \left(l + \frac{3}{2}\right)}{\sqrt{3}}\right)}{5971968 \sqrt[3]{2} \sqrt{3} \left(l + \frac{3}{2}\right)^4} + \frac{1560503 \pi  \left(\omega R-\frac{\sqrt[3]{2} \left(l + \frac{3}{2}\right)}{\sqrt{3}}\right)^2}{71663616 \left(2^{2/3} \left(l + \frac{3}{2}\right)^5\right)}+\cdots .
\end{split}
\eeq
Then the first condition of (\ref{eq:asymQNM}) can be expressed as,
\beq
\label{eq:Condexp}
 2\pi i\lr{n + \frac{1}{2}} = \Pi_{\gamma_{23}}^{(0)} + \Pi_{\gamma_{23}}^{(2)} + \Pi_{\gamma_{23}}^{(4)} + \cdots = \sum_{m=0}^{\infty}c_m\lr{\omega R - \frac{\sqrt[3]{2} \left(l + \frac{3}{2}\right)}{\sqrt{3}}}^m.
\eeq
Applying the Lagrange inversion theorem to (\ref{eq:Condexp}), we obtain an analytic form of the QNMs for $-\frac{\pi}{6} \leq \arg(\omega R) \leq 0$,
\beq
\label{eq:ANQNM}
\hspace*{-0.4in}
 \omega_n R = \frac{\sqrt[3]{2} \left(l + \frac{3}{2}\right)}{\sqrt{3}} + \frac{2\pi i\lr{n + \frac{1}{2}}-c_0}{c_1} - \frac{c_2 (2\pi i\lr{n + \frac{1}{2}}-c_0)^2}{c_1^3} + \cdots .
\eeq
Table.\ref{tab:NumQNMs} shows some numerical values of the QNMs calculated by the first condition in (\ref{eq:exactQNM}) and (\ref{eq:ANQNM}), where we have used up to 4th-order of the WKB period and 8-th order of (\ref{eq:ANQNM}).
\par
To obtain an analytic form of the QNMs for $\arg(\omega R) < -\frac{\pi}{6}$ as (\ref{eq:ANQNM}), we also need to expand (\ref{eq:13int}). But the expansion of (\ref{eq:13int}) at $\omega R = \frac{\sqrt[3]{2}\left(l + \frac{3}{2}\right)}{\sqrt{3}}$ has the logarithm term $\log\left(\omega R - \frac{\sqrt[3]{2}\left(l + \frac{3}{2}\right)}{\sqrt{3}}\right)$ and therefore we cannot compute the inverse series. We need another computational method to obtain an analytic form for $\arg(\omega R) < -\frac{\pi}{6}$.

\begin{table}[t]
\begin{center}
\hspace*{-0.4in} 
\begin{tabular}{c|cccc} \hline
$n$ & $l$ & (\ref{eq:exactQNM}) & (\ref{eq:ANQNM}) & $-\frac{\pi}{6} \leq \arg(\omega R)$  \\ \hline\hline
0 & 0 & $1.057134 - 0.365624i$ & $\bold{\underline{1.0571}}03 - \bold{\underline{0.365624}}i$ & True \\
& 1 & $1.798272 - 0.364387i$ & $\bold{\underline{1.79827}}0 - \bold{\underline{0.364387}}i$ & Ture \\
& 2 & $2.531500 - 0.364053i$ & $\bold{\underline{2.531500}} - \bold{\underline{0.364053}}i$ & True \\ \hline\hline
1 & 1 & $1.535914 - 1.086901i$ & $\bold{\underline{1.53}}6070 - \bold{\underline{1.08}}7184i$ & False \\
& 2 & $2.351632 - 1.089620i$ & $\bold{\underline{2.35163}}8 - \bold{\underline{1.0896}}38i$ & True \\ \hline
\end{tabular}
\caption{Numerical results of the first condition in (\ref{eq:exactQNM}) and (\ref{eq:ANQNM}). The fifth column indicates which the numerical values of the third and fourth columns satisfy $-\frac{\pi}{6} \leq \arg(\omega R)$ or not. The result of  (\ref{eq:ANQNM}) at $n = 1,\ l = 1$, which does not satisfy $-\frac{\pi}{6} \leq \arg(\omega R)$, has lower precision than the other results.}
\label{tab:NumQNMs}
\end{center}
\end{table}

\par
\vspace{0.2in}
\par
In the large $n$ limit, which corresponds to $|\omega R| \gg (l+\frac{3}{2})$ limit, the leading order approximation is valid,
\beq
\label{eq:largenC}
\left\{
\begin{array}{l}
\Pi_{\gamma_{23}}^{(0)} = 2\pi i\left(n + \frac{1}{2}\right)\ \ \ \ \ \ \ \ \ \ \ \ \ \ \ \ \ \ \ \ \ \ \ \ \left(-\frac{\pi}{3} < \arg(\omega R) < 0\right), \\
\Pi_{\gamma_{23}}^{(0)} + \log\left(1 + e^{\Pi_{\gamma_{12}}^{(0)}}\right) = 2\pi i\left(n + \frac{1}{2}\right)\ \ \left(-\frac{\pi}{2} \leq \arg(\omega R) < -\frac{\pi}{3}\right).
\end{array}
\right.
\eeq
In addition, we can neglect the exponentially small term $e^{\Pi_{\gamma_{12}}^{(0)}}$ as we neglect the tunnel effect at the large energy classical limit in quantum mechanics. Then two conditions in (\ref{eq:largenC}) reduce to a single condition,
\beq
\label{eq:ClassicalBS}
 \Pi_{\gamma_{23}}^{(0)} = 2\pi i\left(n + \frac{1}{2}\right).
\eeq
At $|\omega R| \gg (l+\frac{3}{2})$, the potential function $Q_0(z)$ is approximated as $Q_0(z) = -(\omega R)^2\frac{z^3 + 1}{z^3}$. The solutions to $Q_0(z) = 0$ then become $z_1 = -1$, $z_2 = e^{-i\frac{\pi}{3}}$, $z_3 = e^{i\frac{\pi}{3}}$. Now $\Pi_{\gamma_{23}}^{(0)}$ can be expressed as
\beq
\label{eq:largen23}
 \Pi_{\gamma_{23}}^{(0)} = 2i\omega R\int_{e^{-i\frac{\pi}{3}}}^{e^{i\frac{\pi}{3}}}\sqrt{\frac{z^3+1}{z^3}}dz = -\frac{\pi\omega R}{\sqrt[3]{2}\sqrt{3}} \ _2F_1\lrk{\frac{5}{6},\frac{5}{6},2,1}.
\eeq
Substituting (\ref{eq:largen23}) into (\ref{eq:ClassicalBS}), we find that the QNM at large $n$ is pure imaginary,
\beq
\label{eq:largenQNM}
 \omega R = -i\frac{2\sqrt[3]{2}\sqrt{3}}{\ _2F_1\lrk{\frac{5}{6},\frac{5}{6},2,1}}\left(n + \frac{1}{2}\right).
\eeq
In some black hole metrics, it is known that the real part of the QNMs at large $n$ is proportional to the entropy of the black hole (e.g. \cite{gr-qc/0212096, gr-qc/0307020, gr-qc/0301173, 2207.09961}). The entropy of the extremal M5-brane metric (\ref{eq:M5metric}) is zero because it has no event horizon. It is consistent with the real part of (\ref{eq:largenQNM}) is zero.

\section{Summary and discussions}
\label{sec:SandD}
In this paper we have applied the exact WKB analysis to the QNMs spectral problem of the massless scalar perturbation to the extremal M5-brane metric. We have solved the connection problem of the Borel resummed WKB solutions and obtained two exact conditions of the QNMs expressed by the Borel resummed WKB periods. The exact conditions are connected by the Delabaere-Pham formula and clarify the non-perturbative property of the QNMs. In the computation of the QNMs, we have found the Seiberg-Witten/gravity correspondence for the extremal M5-branes to compute the WKB periods. We have calculated the QNMs by using the Borel-Pad\'{e} approximation numerically. We have also computed analytic expressions of the QNMs in some regions.
\par
There are many directions for future work. First, we are interested in applying the exact WKB analysis to other black hole metrics and perturbations. Actually, following the M5-branes, we have already done some computations for the massless scalar perturbation to the extremal M2-branes case based on \cite{hep-th/9702076}. We have drawn the Stokes lines and observed the saddle reduction like the M5-branes case. But for the M2-branes, the classical contributions of the WKB periods are integrals of a hypergeometric curve and it is difficult to compute them. Another interesting direction is the application to the metrics that were studied in the context of the Seiberg-Witten/gravity correspondence \cite{2006.06111, 2105.04245, 2109.09804}. In these papers, it was conjectured that the QNMs are calculated by only the first condition of (\ref{eq:asymQNM}) with suitable WKB period. The conjectures were numerically verified in some lower modes of the QNMs, but it may happen the saddle reduction as we go to higher modes and we may obtain another condition.
\par
Not only the application to other geometries, there are still some questions in the study of the M5-branes. In this paper, we used the Borel-Pad\'{e} approximation to compute the Borel resummed WKB periods because we need to know an infinite number of the coefficients of the WKB periods in order to compute the Borel resummation according to the definition (\ref{eq:BT}), (\ref{eq:Bsumdef}) and it is impossible. But without knowing the higher order coefficients, we can calculate the Borel resummed WKB periods for the quantum Seiberg-Witten curve of 4-dimensional $\mathcal{N} = 2$ SU(2) supersymmetric QCD with a massless fundamental matter by using the TBA equations \cite{2105.03777}. The result of the TBA equations includes the all-order contribution of the WKB periods and therefore is expected to have higher numerical precision than the Borel-Pad\'{e} approximation. 
\par
It is also important to re-derive the QNMs conditions (\ref{eq:BCcoef})$\sim$(\ref{eq:asymQNM}) in different ways. In the ODE/IM method \cite{2112.11434, 2208.14031}, the QNMs conditions is identified with the Bethe roots condition for the Baxter's $Q$-function, which can be expressed by the WKB periods. The QNMs conditions can also be derived by using two-dimensional CFT technique \cite{2105.04483, 2201.04491, 2206.09437}. The differential equation for the QNMs spectral problem is identified with the semiclassical limit of the differential equation for the conformal block with a degenerate field insertion in the  CFT. The connection problem in the CFT then provides the QNMs conditions, which can also be expressed by the WKB periods. The re-derivations of (\ref{eq:BCcoef})$\sim$(\ref{eq:asymQNM}) reinforce the correctness of our QNMs conditions.

\section*{Acknowledgements}
We would like to thank Katsushi Ito, Shota Fujiwara, Tatsuya Mori, Shuichi Murayama for valuable discussions and comments. We are particularly grateful to Katsushi Ito for the useful advice in the preparation of this paper. This work is supported by JSPS Grants-in-Aid for Research Fellows No.22J15182.

\end{document}